\documentclass[11pt]{elsarticle}
\usepackage[T1]{fontenc}
\usepackage[brazil,english]{babel}

\usepackage[utf8]{inputenc}
\usepackage{amsthm,amsfonts,amsmath,amssymb}
\usepackage[ruled,vlined]{algorithm2e}
\usepackage{pifont}
\usepackage{subfigure} 
\usepackage{graphicx}
\usepackage{psfrag}

\usepackage{xcolor}
\usepackage{array}
\usepackage{units}
\usepackage{mdframed}
\usepackage{colortbl}
\usepackage[dvips]{geometry}
\usepackage{setspace}
\usepackage{multirow}
\usepackage{enumitem}
\usepackage{auto-pst-pdf}
\usepackage{booktabs} 
\usepackage{steinmetz} 
\makeatletter
\newcommand{\mypm}{\mathbin{\mathpalette\@mypm\relax}}
\newcommand{\@mypm}[2]{\ooalign{%
  \raisebox{.1\height}{$#1+$}\cr
  \smash{\raisebox{-.6\height}{$#1-$}}\cr}}
\makeatother

\usepackage{url}
\begin{document}

\doublespacing


\pagenumbering{arabic}

\begin{center}
{\Large\bfseries Optimal Energy System Scheduling Using A Constraint-Aware Reinforcement Learning Algorithm}

\vspace{2mm}
{\bfseries $\text{Hou Shengren}^{a}$, $\text{Pedro P. Vergara}^{a*}$, $\text{Edgar Mauricio Salazar Duque}^b$,  and $\text{Peter Palensky}^a$.}
\end{center}

\noindent {\small ${^a}\text{Intelligent Electrical Power Grids (IEPG)}$ Group, Delft University of Technology, Delft 2628CD, The Netherlands.}

\noindent {\small${^b}\text{Electrical Energy Systems (EES)}$ Group, Eindhoven University of Technology, Eindhoven 5612AE, The Netherlands.}

\noindent {\small emails: h.shengren@tudelft.nl, p.p.vergarabarrios@tudelft.nl, e.m.salazar.duque@tue.nl, p.palensky@tudelft.nl} 

\noindent ${^*}\text{Corresponding author}$


\section*{Abstract}
The massive integration of renewable-based distributed energy resources (DERs) inherently increases the energy system's complexity, especially when it comes to defining its operational schedule. Deep reinforcement learning (DRL) algorithms arise as a promising solution due to their data-driven and model-free features.  However, current DRL algorithms fail to enforce rigorous operational constraints (e.g., power balance, ramping up or down constraints) limiting their implementation in real systems. To overcome this, in this paper, a DRL algorithm (namely MIP-DQN) is proposed, capable of \textit{strictly} enforcing all operational constraints in the action space, ensuring the feasibility of the defined schedule in real-time operation. This is done by leveraging recent optimization advances for deep neural networks (DNNs) that allow their representation as a MIP formulation, enabling further consideration of any action space constraints. Comprehensive numerical simulations show that the proposed algorithm outperforms existing state-of-the-art DRL algorithms, obtaining a lower error when compared with the optimal global solution (upper boundary) obtained after solving a mathematical programming formulation with perfect forecast information; while strictly enforcing all operational constraints (even in unseen test days).

\noindent {\bfseries Keywords:} Energy management systems, distributed energy system, safe reinforcement learning, machine learning, nonlinear programming.

\singlespacing
\begin{mdframed}
\section*{Notation}
\noindent The notation used throughout this paper is reproduced below for reference.
\noindent \emph{Sets}:
\begin{description}[itemsep=0.5ex, labelwidth=2.5cm, leftmargin=1.7cm]
	\item [{\small${\cal G, B, L, V}$}]      Set of (DGs) distributed generators, EESs, Loads and PVs. 
	\item [{\small${\cal S, A}$}]    Set of states, set of actions.
	\item [{\small${\cal T}$}]      Set of time steps
	
 \end{description}

\noindent \emph{Indexes}:
\begin{description}[itemsep=0.5ex, labelwidth=2.5cm, leftmargin=1.7cm]
	\item [{\small$i$}]  DG unit $i\in{\cal{G}}$
	\item[{\small$j$}] ESS $j\in\cal{B}$
	\item[{\small$m$}] PV unit $m\in\cal{V}$
	\item[{\small$k$}] Load demand $m\in\cal{L}$
	\item [{\small$t$}]  Time-step $t\in\cal{T}$
\end{description}

\noindent \emph{Parameters}:
\begin{description}[itemsep=0.5ex, labelwidth=2.5cm, leftmargin=1.7cm]
	\item [{\small$\theta,\theta^{\text{target}},\omega$}]    	   	Parameters for the DNN's $Q_{\theta}$, $Q_{\theta^{\text{target}}}$ and $\pi_{w}$
	\item [{\small$a_{i}, b_{i}, c_{i}$}]    	   	Quadratic, linear and constant parameters associated to the $i$-th DG operation cost
	\item [{\small$\Delta t$}]    		Length discretization of the operational time
	\item [{\small$\gamma$}]    		Discount factor 
	\item [{\small$\overline{P}_{t}^{G}, \underline{P}_i^{G}$}]    	Maximum/minimum generation limit of the DG units
	\item[\small$RU_{i}, RD_{i}$]	Ramping up/ramping down ability of the DG units 
    \item[{\small$\overline{P}_{j}^{B}$~$\underline{P}_{j}^{B}$}]     Maximum/minimum charging/discharging limit of the ESSs
    \item[{\small$\overline{SOC}_{j}^{B}$}]     Maximum SOC of the ESSs
    \item[{\small$\underline{SOC}_{j}^{B}$}]     Minimum SOC of the ESSs
    \item[{\small$E{j}^{B}$}]     Energy capacity of the ESSs
    \item[\small$\overline{P}^{C}$]
    Maximum main network export/import limit
	\item [{\small$\beta$}]    			Electricity sell coefficient
	\item [{\small$\eta_{B}$}] Energy exchange efficiency for ESSs
	\item [{\small$\sigma_{1}, \sigma_{2}$}]   Reward re-scale and constrain penalty coefficients	
	\item[{\small$\rho_{t}$}] Electricity price for time slot $t$
	\item [{\small$P^{V}_{m,t}$}]    		    Active power of PV systems
	\item[{\small$P^{L}_{k,t}$}]	Active power demand 
	\end{description}
\noindent \emph{Continuous Variables}:
\begin{description}[itemsep=0.5ex, labelwidth=2.5cm, leftmargin=1.7cm]
	\item [{\small$P^{G}_{i,t}$}]    		Active power output of DG units
	\item [{\small$P^{B}_{j,t}$}]    			Active	power discharge/charge of ESSs
	\item [{\small$SOC_{j,t}^{B}$}]    			State of charge for ESSs
	\item[{\small$P_{t}^{N}$}]
		Active power exported/imported to/from the main network
	\item[{\small$\Delta P_{t}$}]	Active power unbalance 
\end{description}
\end{mdframed}

\doublespacing
\section{Introduction}
\label{sec:introduction}To reduce the impact of the energy sector on the environment, distributed energy resources (DERs) are being integrated into our energy systems. Such DERs, in the form of renewable-based systems (e.g., PV systems and wind turbines) and small-scale energy storage systems (ESSs), provide more flexibility, enabling a more efficient operation. Nevertheless, these DERs also increase the energy system's complexity, especially when it comes to defining its operational schedule. Moreover, due to their weather-dependent nature, renewable-based DERs inherently increase the energy system's levels of uncertainty, requiring scheduling algorithms capable of providing fast and good-quality, but feasible, solutions~\cite{ZiaElbouchikhi2018}. In the technical literature, two main approaches are available to deal with the optimal scheduling of energy systems; namely, \textit{model-based} and \textit{model-free} approach. A detailed literature review is presented next. 

\subsection{Literature Review}

In general, model-based approaches rely on precise models to build complex mathematical formulations in order to consider the energy system' operational constraints. Depending on how these constraints are modeled, the derived mathematical formulations can be classified as linear, nonlinear programming, or dynamic programming problems~\cite{zambroni_de_souza_microgrids_2019}. In this regard, in~\cite{VergaraLopez2019}, a mixed-integer nonlinear programming (MINLP) formulation is used to determine the optimal operation of an unbalanced three-phase energy system. In order to reduce the complexity of the proposed formulations, linearizations and simplifications are introduced. Similar work has been done in \cite{GiraldoCastrillon2019}. Nevertheless, the model-based nature of these methods requires considerable precision of the built mathematical models, which limits their performance, especially if uncertainty is to be considered. 

Generally, in model-based approaches, uncertainty is modeled either by using a probability distribution function or by leveraging a set of representative scenarios, leading to stochastic or robust mathematical formulations, such as the ones presented in~\cite{MojtabaHajizadeh2019,YousefiHajizadeh2021,VergaraLopezStch2020}. Other approaches, such as the one in~\cite{arroyo2022reinforced}, leverage a rolling time horizon approach to eliminate the forecast error when defining the DERs optimal energy scheduling. To guarantee the feasibility of the defined schedule under various operational scenarios, in~\cite{chen2022robust}, an adjustable two-stage robust optimization framework is proposed, solving simultaneously a day-ahead scheduling and real-time regulation problem of an integrated energy system. In~\cite{su2022energy}, a chance-constrained programming model is proposed to schedule an active distribution network incorporating office buildings. Nevertheless, modeling the probability distribution of uncertain data is challenging, while using a large number of scenarios may cause a computational burden. Therefore, although capable of providing good quality solutions, existing model-based approaches are not adequate for handling the increased uncertainty level of renewable-based energy systems, as their performance and efficiency mainly depend on the accuracy of the used models and their approximations. Moreover, the computational complexity of these methods increases dramatically with the system size, imposing scalability and convergence challenges.

To overcome this, model-free approaches have been introduced as an alternative solution. The most promising approach is based on the use of reinforcement learning (RL)~\cite{ChenQu2021}, modeling the decision-making problem as a Markov Decision Process (MDP). One of the most interesting features of RL algorithms is that they can learn any system's dynamics by interaction, providing good-quality solutions guided by a reward value used as a performance indicator~\cite{SuttonBarto2018}. Recently, deep reinforcement learning (DRL) algorithms have shown good performance when solving MDPs in energy systems tasks~\cite{VazquezRey2020}, ranging from, home energy management~\cite{nakabi_deep_2021}, microgrid dispatch~\cite{ji_real-time_2019}, voltage regulation~\cite{wang2021multi}, and electricity network operation~\cite{kelly2020reinforcement}. Other applications include, for instance, a standarized DRL approaches for demand response in smart buildings~\cite{VazquezRey2020}, and learning to solve fast optimal power flow problems using DRL algorithms, specifically  the proximal policy optimization (PPO) algorithm and imitation learning~\cite{zhou2020data}. In~\cite{pinto2021data}, a performance comparison of the soft actor-critic (SAC) algorithm with a rule-based control method on the surrogate simulation model developed by~\cite{VazquezRey2020}, is presented. In~\cite{wang2021multi}, the voltage regulation problem of a distribution network is first modeled as a partial-observable MDP, and then multi-agent DRL algorithms are leveraged to execute the optimal solutions. In~\cite{heidari2022reinforcement}, a DRL approach-based proactive operation framework is proposed to model the stochastic behavior and uncertainty of solar energy for residential buildings. In~\cite{liu2021deep}, a DRL algorithm is developed to solve a stochastic energy management problem considering power flow constraints, resulting in an optimal policy that minimizes total operational cost (although operational constraints are disregarded).

Different from the energy-related MDPs presented above, the operational schedule of DERs within an energy system must enforce a rigorous set of operational constraints to ensure a reliable and safe operation, e.g.,~generation and consumption must always be balanced during real-time operation, ramping-up and down constraints, etc. Nevertheless, current DRL algorithms lack of safety guarantees~\cite{MassianiPierreFrancois2021}, as these constraints cannot be directly imposed in the algorithm's formulation. Different strategies to indirectly enforce operational constraints have been proposed to overcome this. In~\cite{zhou_combined_2020}, a DG unit is set as a slack bus with unlimited generation capacity, avoiding unbalance by the outputs of the generators controlled by DRL agents. In~\cite{YingXu2021}, a penalty term is added to the reward function to guide the learning process aiming to reduce operating costs while enforcing power balance. A similar penalty approach has been used to enforce voltage magnitude constraints in case the electricity network operation is considered. For instance,~\cite{vergara_optimal_2019} modeled the dispatch of PV inverters as an MDP, and built a decentralized dispatch framework penalizing RL agents when actions lead to voltage violations. In research~\cite{mauricio2022eligibility}, an on-policy RL algorithm with eligibility traces is developed to dispatch the energy storage system to minimize the cost and regulate voltage magnitudes. A similar work is presented in~\cite{liu2018distributed}. In~\cite{du2022deep}, a service assistant restoration problem is modeled as MDP. Then, imitation learning is employed as expert demonstrations enabling a deep deterministic policy gradient (DDPG) agent learn a safe policy for online implementation. In~\cite{qiu2022coordination}, a double auction market-based coordination framework is proposed to schedule the energy trading between multi-energy microgrids. Multi-agent twin delayed deep deterministic algorithm (TD3) is used to solve the formulated problem, while a large penalty is imposed on the reward function to reduce the energy unbalance. In~\cite{yi2022improved}, the SAC algorithm is leveraged to control a virtual power plant to provide frequency regulation services, penalizing any frequency deviation. Nevertheless, although these strategies may enforce operational constraints during training, they are either based on nonpractical assumptions or fail to guarantee the feasibility of the defined operating schedule in real-time, especially in cases of large peak consumption or renewable-based generation~\cite{shengren2022performance}. 

Strategies based on safe RL have also been proposed to directly enforce operational constraints, exploiting results from different research areas, such as robot manipulation~\cite{encode_knowledge,comprehensive_survey_SRL_review}. In~\cite{srl_ed}, an action projection layer is implemented, correcting the action defined by the DRL algorithm via a projection operator. Unfortunately, this projection operator degrades the DRL algorithm's performance, as shown in~\cite{srl_projection_performance_test}. In~\cite{qiu2022safe}, safe DDPG is used for real-time automatic control of a smart hub, while a safety net is used to estimate the feasibility of decided actions. A similar strategy is proposed in~\cite{DIP_QL}, in which the action proposed by the DRL algorithm is used as starting point to solve a mathematical programming formulation, ensuring constraints compliance. In \cite{cpo_distirbution_network}, a constrained policy gradient approach is proposed, updating the parameters of the DNN model in the direction that minimizes the power unbalance. In~\cite{li2019constrained}, the same approach is used to solve an EVs coordination problem. This policy optimization approach allows the DRL algorithm to provide a probabilistic notion of safety. Nevertheless, feasibility is paramount in energy systems operation, and it should be certifiable. In this regard, enforcing operational constraints during the online scheduling stage is a critical challenge for DRL algorithms and it must be addressed in order to enable their wide adoption in real systems. A summary of the discussed research literature is presented in Table~\ref{summary_literature_review}. The openness and free online availability of the algorithms discussed here are also highlighted in Table~\ref{summary_literature_review}.   

\begin{table}
\centering
\caption{Summary of research literature for DRL algorithms and constraint enforcing approaches.}
\label{summary_literature_review}
\scalebox{0.63}{
\begin{tabular}{llllll} 
\hline
\textbf{Work}                               & \textbf{Research Problem}                         & \textbf{Constraint Enforcing}                                                 & \textbf{Advantages}                                                                             & \textbf{Disadvantages}                                                                                                       & \textbf{Open-access~}  \\ 
\hline
\cite{VazquezRey2020}             & Residential building energy schedule      & \multirow{3}{*}{Constraints disregarded~~}                                          & \multirow{3}{*}{Simple~}                                                                        & \multirow{3}{*}{Not realistic~}                                                                                              & Yes                    \\
\cite{nakabi_deep_2021}         & Microgrid operation~                      &                                                                                &                                                                                                 &                                                                                                                              & Yes~                   \\ 

\cite{heidari2022reinforcement}   & Residential buildings energy schedule      &                                                                                &                                                                                                 &                                                                                                                              & No                   \\ \hline
\cite{ji_real-time_2019}        & Microgrid operation~                      & \multirow{9}{*}{Penalty function~}                                             & \multirow{9}{*}{Easy to implement~}                                                             & \multirow{9}{*}{No constraint guarantee~}                                                                                               & No                     \\
\cite{wang2021multi}              & Voltage regulation~                       &                                                                                &                                                                                                 &                                                                                                                              & Yes                    \\
\cite{zhou2020data}               & Optimal power flow~                       &                                                                                &                                                                                                 &                                                                                                                              & No                     \\
\cite{liu2021deep}                & Energy dispatch~                        &                                                                                &                                                                                                 &                                                                                                                              & No                    \\ 
\cite{YingXu2021}                 & Energy dispatch~                        &                                                                                &                                                                                                 &                                                                                                                              & No                     \\ 
\cite{shengren2022performance}    & Optimal energy system scheduling~         &                                                                                &                                                                                                 &                                                                                                                              & Yes~                   \\
\cite{vergara_optimal_2019}     & PV-inverter voltage regulation~     &                                                                                &                                                                                                 &                                                                                                                              & No                     \\
\cite{mauricio2022eligibility}    & Battery schedule and voltage regulation &                                                                                &                                                                                                 &                                                                                                                              & No                     \\
\cite{qiu2022coordination}        & Energy trading between microgrids        &                                                                                &                                                                                                 &                                                                                                                              & No                     \\ \hline
\cite{du2022deep}                 & Restoration services         & \begin{tabular}[c]{@{}l@{}}Imitation learning and \\penalty function\end{tabular} & \begin{tabular}[c]{@{}l@{}}Accelerating training speed \\Improve the performance~\end{tabular} & No constraint guarantee~~                                                                                                               & No                  
\\ \hline
\cite{zhou_combined_2020}       & Energy Management~                        & Unlimited slack bus~                                                   & Simple~                                                                                         & Not realistic~                                                                                                               & Yes                    
\\ \hline
\cite{srl_ed}                    & Energy management~                        & Safe layer~                                                                    & \multirow{3}{*}{Guarantee the feasibility}                                                      & \multirow{3}{*}{\begin{tabular}[c]{@{}l@{}}Performance deterioration \\Not fully model-free\end{tabular}}                   & \multirow{3}{*}{No}    \\
\cite{qiu2022safe}               & Energy hub trading~                       & \begin{tabular}[c]{@{}l@{}}Gaussian process\\Safe layer\end{tabular}          &                                                                                                 &                                                                                                                              &               \\ 
\cite{DIP_QL}                    & Microgrid operation~                     & Action projection~                                                             &                                                                                                 &                                                                                                                              &                        \\ \hline
\cite{cpo_distirbution_network} & Distribution network operation            & \multirow{2}{*}{Constrained policy optimization~}                              & \multirow{2}{*}{Probabilistic guarantee feasibility~}                                           & \multirow{2}{*}{\begin{tabular}[c]{@{}l@{}} No constraint guarantee\\Higher computation time\end{tabular}} & \multirow{2}{*}{No}    \\
\cite{li2019constrained}         & EV management~                            &                                                                                &                                                                                                 &                                                                                                                             &                        \\
\cline{1-6}
\end{tabular}}
\end{table}

\subsection{Contributions}
To overcome the above-discussed limitations, this paper proposes a DRL algorithm (namely MIP-DQP) to define the optimal schedule of a renewable-based energy system, capable of \textit{strictly} enforcing all the operational constraints in the action space, ensuring the feasibility of the defined scheduled in real-time operation. To do this, we used recent optimization advances for DNNs that allow their representation as a mixed-integer linear (MIP) formulation, enabling further consideration of any action space constraints. Such approaches have been also employed in the context of feature visualization and adversarial machine learning~\cite{fischettiJo2018}. The performance of the proposed algorithm has been compared with other state-of-the-art DRL algorithms available in the literature, including DDPG, PPO, SAC, and TD3 algorithms~\cite{ChenQu2021}, to show its effectiveness. A comparison with the optimal global solution is also presented, obtained by solving the energy system scheduling problem as a mathematical programming formulation considering full knowledge of future information (i.e., consumption, dynamic prices, and renewable-based generation). The main contributions of this paper are as follows: 
\begin{itemize}
    \item A value-based DRL algorithm to solve the energy system scheduling problem is proposed, capable of dealing with continuous action spaces. Different from other actor-critic DRL algorithms (e.g., DDPG, PPO, and TD3~\cite{ChenQu2021}), we make use of the action-value function approximated using a DNN, while discarding the policy model learning used during exploration. 
    \item An innovative online execution approach that guarantees that the proposed DRL algorithm \textit{strictly} meets all operational conditions in the action space (e.g., the power balance constraint), even in unseen test data, is also proposed. This is done by leveraging new optimization results from DNNs that allow their representation as a MIP formulation, enabling further consideration of any action space constraints.  
\end{itemize}

The rest of this paper is organized as follows. In Section 2, the optimal energy system scheduling problem is formulated. Then, in Section 3, the formulated problem is modeled as MDP while the proposed MIP-DQN algorithm is illustrated and used to solve the optimal energy system scheduling problem in Section 4. Simulation tests are presented, analyzed and discussed in Section 5, while final conclusions are presented in Section 6.

\vspace{-2mm}
\section{Mathematical Programming Formulation of the Energy Systems Scheduling Problem}\label{miqp_formulation}

\begin{figure}[t]
   \centering
    \psfrag{A1}[][][0.8]{$P^{V}_{m,t}$}
    \psfrag{A2}[][][0.8]{$P^{L}_{k,t}$}
    \psfrag{A3}[][][0.8]{$P^{G}_{i,t}$}
    \psfrag{A4}[][][0.8]{$P^{B}_{j,t}$}
    \psfrag{A5}[][][0.8]{$P^{N}_{t}$}
    \includegraphics[width=0.5\columnwidth]{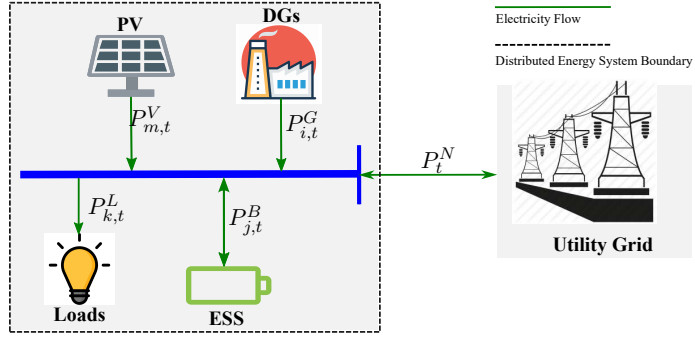}
   \caption{Illustration of the considered energy system structure composed of various DERs, such as solar photovoltaic (PV), ESSs, DGs, and loads.}
   \label{fig_illustration}
\end{figure}

The structure of the considered energy system is shown in Fig.~\ref{fig_illustration}, including various DERs, such as solar photovoltaic (PV), ESSs, DGs, and loads, while a connection to the utility grid is leveraged to address a demand surplus or shortage problem. For tractable analysis, we assume the day-ahead market where the electricity price of each hour is revealed beforehand. For the energy system in Fig.~\ref{fig_illustration}, the optimal energy system scheduling problem can be modeled by the nonlinear programming (NLP) formulation described by \eqref{eq_obj}-\eqref{eq_export_cons}. The objective function in \eqref{eq_obj} aims at minimizing the operating cost for the whole time horizon ${\cal T}$, comprising the operating cost of the DG units, as presented in \eqref{eq_gen_cost}, and the cost of buying/selling electricity from/to the main network, as in \eqref{eq_power_exchange_cost}. Given the output power of DG units $P_{i,t}^G$, the operating cost can be estimated by using a quadratic function as in \eqref{eq_gen_cost}. The transaction cost between the energy system and the network is settled according to Time-of-Use (ToU) prices, in which it is assumed that selling prices are lower than the purchasing prices. In \eqref{eq_power_exchange_cost}, $\rho_{t}$ is the ToU price at time slot $t$, while $P_{t}^{N}$ refers to the exported/imported power transaction to/from the network.

\vspace{-2mm}
\begin{equation}\label{eq_obj}
\min_{\substack{P^{G}_{i,t}, P_{j,t}^{B}}} \left\{  \sum_{t\in\cal{T}}\sum_{i\in\cal{G}} \left[ C_{i,t}^{G}(\cdot)+C_{t}^{E}(\cdot)\right]\Delta t \right\},
\end{equation}

\vspace{-2mm}
\begin{equation}\label{eq_gen_cost}
C_{i,t}^{G}=a_{i}\left(P_{i,t}^{G}\right)^{2}+b_{i} P_{i,t}^{G}+c_{i},  \quad \forall i \in \cal{G}.
\end{equation}

\vspace{-2mm}
\begin{equation}\label{eq_power_exchange_cost}
C_{t}^{E} = \begin{cases}
		\rho_{t}P_{t}^{N}	  &  \quad P_{t}^{N}>0,\\
		\beta\rho_t P_{t}^{N} & \quad P_{t}^{N}<0.\\
	\end{cases}
\end{equation}

Subject to: 
\vspace{-2mm}
\begin{flalign}
& \sum_{i\in\cal{G}}P_{i,t}^{G} +\sum_{m\in\cal{V}}P_{m,t}^{V}+P_{t}^{N}+\sum_{j \in\cal{B}}P_{j,t}^{B} = \sum_{k \in \cal{L}}P_{k,t}^{L}
& \forall t \in \cal{T}
&\label{eq_balance}\\
&\underline{P}^{G}_{i} \leq P^{ G}_{i,t} \leq \overline{P}^{G}_{i}
& \forall i \in {\cal G}, \forall t \in {\cal{T}} 
& \label{eq_max_min_output_constrain} \\
&P_{i,t}^{G}-P_{i,t-1}^{G}\leq RU_{i}
& \forall i \in {\cal G}, \forall t \in {\cal{T}} 
& \label{eq_ramping_up_constrain} \\
&P_{i,t}^{G}-P_{i,t+1}^{G}\leq RD_{i}
& \forall i \in {\cal G}, \forall t \in {\cal{T}} 
& \label{eq_ramping_down_constrain} \\
&-\underline{P}_{j}^{B}\leq P_{j,t}^{B}\leq \overline{P}_{j}^{B}
&  \forall j \in {\cal{B}}, \forall t \in {\cal{T}}
& \label{eq_char_disc_cons} \\
&SOC_{j,t}^{B}=SOC_{j,t-1}^{B} + \eta_{B}P_{j,t}^{B}\Delta t/E^{B}_{j}
&\forall j \in {\cal{B}}, \forall t \in {\cal{T}}
&\label{eq_SOC_cha} \\
&\underline{SOC}_{j}^{B}\leq SOC_{j,t}^{B}\leq\overline{SOC}_{j}^{B}
&\forall j \in {\cal{B}}, \forall t \in {\cal{T}}
&\label{eq_SOC_cons} \\
&-\overline{P}^{C}\leq P_{t}^{N}\leq \overline{P}^{C}
&\forall t \in {\cal{T}}
& \label{eq_export_cons} 
\end{flalign}

Expression \eqref{eq_balance} defines the power balance constraint. Expression \eqref{eq_max_min_output_constrain} defines the DG units generation power limits while \eqref{eq_ramping_up_constrain} and \eqref{eq_ramping_down_constrain} enforce the DG unit's ramping up and down constraints, respectively. Energy storage systems (ESSs) are modeled using \eqref{eq_char_disc_cons}-\eqref{eq_SOC_cons}. In this model, the operation cost of ESSs is not considered, while ESSs are allowed to schedule their discharge and charge power in advance. Expression \eqref{eq_char_disc_cons} defines the charging and discharging power limits, while expression \eqref{eq_SOC_cha} models the state of charge (SOC) as a function of the charging and discharging power. Expression in \eqref{eq_SOC_cons} limits the energy stored in the ESSs, avoiding the impacts caused by over-charging and over-discharging. Finally, the main network export/import power limit is modeled by the expression in~\eqref{eq_export_cons}. Notice that in order to solve the mathematical formulation described by \eqref{eq_obj}-\eqref{eq_export_cons}, \textit{full knowledge} of future information (e.g., renewable-based generation, consumption and dynamic prices) is required, for instance, provided via a forecasting algorithm. The proposed DRL algorithm is able to provide good-quality solutions with only current information, as shown later. Next, the MDP formulation of the optimal scheduling problem is presented. 

\vspace{-2mm}
\section{MDP Formulation \& Value-Based DRL}\label{mdp_formulation}
The above-presented decision-making problem can be modelled as a finite MDP, represented by a 5-tuple $(\cal{S},\cal{A},\cal{P},\cal{R},\gamma)$, where $\cal{S}$ represents the set of system states, $\cal{A}$ the set of actions, $\cal{P}$ the state transition probability function, $\cal{R}$ the reward function, and $\gamma$ a discount factor. In this formulation, the energy system operator can be modeled as an RL agent. The state information provides an important basis for the operator to dispatch units. We define a state at time $t$ as $s_{t}= (P_{t}^{V},P_{t}^{L},P_{t-1}^{G},SOC_{t}), \quad s_{t}\in \mathcal{S}$, while the actions, defining the scheduling of the DG units and the ESSs, as $
a_{t}= (P^{G}_{i,t},P_{t}^{B}), \quad a_{t}\in \mathcal{A}$. Notice that the RL agent does not directly control the transaction between the energy system and the main network (i.e., $P_{t}^{N}$). Instead, after any action is executed, power is exported/imported from the main network to maintain the power balance. Nevertheless, a maximum power capacity constraint exists and must be enforced i.e.,~\eqref{eq_export_cons}. Notice that if the maximum export/import limits are defined to be a low value (as done in this paper), in most cases, the power balance constraint will not be automatically met.

Given the state $s_{t}$ and action $a_{t}$ at time step $t$, the energy system transit to the next state $s_{t+1}$ defined by the next transition probability function
\vspace{-2mm}
\begin{equation}
p(S_{t+1},R_{t}|S_t,A_t)= \\ \operatorname{Pr}\left\{S_{t+1}=s_{t+1}, R_{t}=r_{t} \mid S_{t}=s_{t}, A_{t}=a_t\right\},
\end{equation}
which models the energy system's dynamics. In model-based algorithms, the uncertainty is predicted by a determined value or sampling from a prior probability distribution. In contrast, DRL algorithms are a model-free approach, capable of learning such dynamics from interactions. To guide learning, a reward $r_{t}$ must be provided by the environment in order for the RL agent to quantify the goodness of any action taken. In the energy system scheduling problem, the reward function ${\cal R}(s_t,a_t)$ should guide the RL agent to take actions that minimize the total operating cost, while enforcing the power balance constraint. This can be done by using the reward function
\vspace{-2mm}
\begin{equation}\label{eq_reward}
{\cal R}_{t}\left(s_{t}, a_{t}\right)= r_t = -\sigma_{1}\left[\sum_{i\in\cal{G}} \left(C^{G}_{i,t}+C^{E}_{t}\right)\right] -\sigma_{2} \Delta P_t, \forall t \in {\cal{T}},
\end{equation}
in which $\Delta P_t$ corresponds to the power unbalance at time-step~$t$, defined as,

\begin{equation}\label{eq_unb}
\Delta P_t=\left\lvert \sum_{i\in\cal{G}}P_{i,t}^{G}+\sum_{m\in\cal{V}}P_{m,t}^{V}+P_{t}^{N}+\sum_{j \in\cal{B}}P_{j,t}^{B}-\sum_{k \in \cal{L}}P_{k,t}^{L}\right\rvert.
\end{equation}

\noindent In \eqref{eq_reward}, $\sigma_1$ and $\sigma_2$ are used to control the order of magnitude and the trade-off between the operating cost minimization and the penalty incurred in case of power unbalance. The procedure used to solve the proposed MDP using value-based RL algorithms is presented next.

\subsection{DRL Value-Based Algorithms}
Define $Q_{\pi}(S_t,A_t)$ as the action-value function that estimates the \textit{expected cumulative reward} given that action $a_t$ is taken at state $s_t$ and following policy $\pi(\cdot)$ after that. The action-value function $Q_{\pi}(S_t,A_t)$ can be expressed recursively as~\cite{SuttonBarto2018},
\begin{equation}\label{eq_recursive_q}
    Q_{\pi}(S_t,A_t) =  \mathbb{E}_{\pi}\left[r_t + \gamma Q_{\pi}(s_{t+1},a_{t+1})|S_t=s_t,A_t=a_t\right].
\end{equation}

Bellman's principle of optimality states that the optimal action-value function for an MDP has the recursive expression
\begin{equation}\label{eq_recursive_q_2}
    Q_{\pi}^{*}(S_t,A_t) =  \mathbb{E}_{\pi}\left[r_t \right. \\ + \left. \gamma \max_{\substack{a_{t+1} \in {\cal A}}} Q_{\pi}^{*}(s_{t+1},a_{t+1})|S_t=s_t,A_t=a_t\right],
\end{equation}
which solution can be obtained by using a Temporal Difference (TD) algorithm~\cite{qlearning_watkins}, which solves the following update rule iteratively.
\begin{equation}\label{eq_recursive_q_approx}
    \hat{Q}(S_t,A_t) \doteq \hat{Q}(S_t,A_t)  + \\ \alpha \left[r_{t} + \gamma \max_{\substack{a_{t+1} \in {\cal A}}} \hat{Q}(s_{t+1},a_{t+1}) - \hat{Q}(S_t,A_t)\right],
\end{equation}
in which $\hat{Q}(\cdot)$ corresponds to a function approximator used to represent $Q_{\pi}^{*}(\cdot)$ and $\alpha \in (0,1]$ is a learning rate. Once a good quality representation of $Q_{\pi}^{*}(\cdot)$ is obtained via $\hat{Q}(\cdot)$, at time step $t$ and state $s_t$, optimal actions $a_{t}$ can be sampled from the optimal policy, i.e., $a_t \sim \pi^{*}(s_t) $, obtained as 
\begin{equation}\label{eq_optimal_action}
\pi^{*}(S_t) = \max_{\substack{a \in {\cal A}}} \hat{Q}(S_t=s_t,a).
\end{equation}

\noindent For continuous state and action spaces, the optimal action-value function $Q_{\pi}^{*}(\cdot)$ can be approximated using a DNN i.e., $\hat{Q}(\cdot)=Q_{\theta}(\cdot)$ with parameters $\theta$, leading to an algorithm known as deep Q-networks (DQNs)~\cite{mnih2015DQN}. In this case, the iterative procedure shown in \eqref{eq_recursive_q_approx} can be seen as a regression problem whose objective is to estimate the DNN's parameters $\theta$ via stochastic gradient ascent. In DQNs, the $Q_{\theta}$ is updated using the value $r_t + \gamma \max_{\substack{a \in {\cal A}}} Q_{\theta^{\textit{target}}} (s_t,a)$, where $Q_{\theta^{\text{target}}} $ is a \textit{target} Q-function\footnote{i.e., a copy of model $Q_{\theta}$ which parameters are updated less frequently. This procedure helps to stabilize learning within the DRL algorithm. For a more detailed explanation, see~\cite{Ryu2020CAQL}.}. Under this value definition, parameters $\theta$ can be obtained minimizing a loss function over mini-batches $B$ of past data $\left\{(s_t,a_t,r_t,s_{t+1})\right\}_{i=1}^{|B|}$. In this case, the loss definition used to train the DQN is based on the mean squared Bellman error, defined as\footnote{For a more detailed derivation of the loss function in \eqref{eq_Q_update}, see~\cite{Ryu2020CAQL}.}
\begin{equation}
    \label{eq_Q_update}
    \min _{\theta} \sum_{i=1}^{|B|}\left(r_{t,i}+\gamma Q_{\theta^{\text {target}}}\left(s_{t+1,i}, \arg \max _{a} Q_{\theta}\left(s_{t+1,i}, a\right)\right) - Q_{\theta}\left(s_{t+1,i}, a_{t,i}\right)\right)^{2}.
\end{equation}

\noindent Notice that in  continuous action spaces, the procedure used in \eqref{eq_optimal_action} to sample actions from the action-value function $Q_{\theta}$ is not feasible since an exhaustive action enumeration (i.e., the Max-Q problem) is not possible. Moreover, in \eqref{eq_optimal_action} actions constraints are completely disregarded. To overcome this, we combine value-based DRL algorithms with mixed-integer programming, as explained next. 

\vspace{-2mm}
\section{Proposed MIP-DQN Algorithm}
The proposed DRL algorithm is named MIP-DQN and is defined through two main procedures: training and deployment (or online execution). The main objective of the training procedure is to estimate the parameters $\theta$ of the DNN used to approximate the action-value function $Q_{\theta}$; whereas during deployment, the obtained function $Q_{\theta}$ is used to take actions to directly operate assets within the energy system. Both procedures are explained in detail below. 

\vspace{-2mm}
\subsection{Training Procedure}
The training process developed for the MIP-DQN algorithm is described in Algorithm~1. This process starts by randomly initializing the parameters of the DNN functions $Q_{\theta}$, $Q_{\theta^{\text{target}}}$. Then, interactions with a model of the energy system take place. In traditional valued-based RL algorithms, exploration is done by sampling actions from the current estimate of the action-value function $Q_{\theta}$. However, and as explained before, sampling actions from $Q_{\theta}$ following \eqref{eq_optimal_action} is not a feasible procedure in continuous action spaces. Instead, we propose to use a parameterized deterministic optimal policy $\pi_{\omega}$, which is also approximated using a DNN model and randomly initialized. Similar to other works~\cite{Ryu2020CAQL,actor_expert}, the policy function $\pi_\omega$, the action-value functions $Q_\theta$ and $Q_{\theta^{\text{target}}}$, will be jointly approximated. 

Within one epoch, for each time step $t$, a transition tuple of the form $(s_t,a_t,r_t,s_{t+1})$ is collected and store in a replay buffer $R$. Then, a subset $B$ of these samples is selected and used to update the parameters of functions $Q_{\theta}$, $Q_{\theta^{\text{target}}}$ and $\pi_\omega$ as shown in Algorithm~1. This procedure is iteratively done until a maximum number of epochs is reached. 

Different from other DRL algorithms, such as DDPG and PPO, after training, we make use of the action-value function $Q_{\theta}$ and discard the approximated policy $\pi_\omega$. Moreover, it is critical to notice that the power balance constraint is only enforced via the penalty added to the reward function in \eqref{eq_reward}. Thus, it is expected that at the end of the training procedure, such equality constraint is not \textit{strictly} met. The procedure used to enforce constraints is developed for the deployment or online execution, as explained next. 

\begin{algorithm}[t]\label{algorithm1}
	\caption{Training procedure for MIP-DQN}
	Define the maximum training epochs $T$, episode length $L$.
	Initialize parameters of functions $Q_{\theta}$, $Q_{\theta^{\text{target}}}$, and $\pi_\omega$;
	Initialize reply buffer $R$. \;
	
	\For{$t=1$ \KwTo $T$}
	{Sample an initial state $s_0$ from the initial distribution
	    
	    \For {$l=1$ \KwTo $L$}
	    {
		Sample an action with exploration noise $a_t \sim \pi_\omega(s_t)+\epsilon$,
		$\epsilon \sim \mathcal{N}(0, \sigma)$ and observe reward $r_t$ and new state $s_{t+1}$. \;
	    \par
	    Store transition tuple $\left(s_t, a_t, r_t, s_{t+1}\right)$ in $R$.\;}
		\par
		Sample a random mini-batch of $|B|$ transitions $\left(s_t, a_t, r_t, s_{t+1}\right)$ from $R$.\;
		\par
		Update the Q-function parameters by using \eqref{eq_Q_update}.\;
		\par
		Update the execution policy function parameters by using $\omega\leftarrow \omega+\nabla_{\omega} \frac{1}{|B|} \sum_{s \in B} Q_{\theta}\left(s, \pi_{\omega}(s)\right)$.
		\par
		Update the target-Q function parameters:
		\begin{equation*}
            \theta^{\text{target}} \leftarrow \tau \theta+(1-\tau) \theta^{\text{target}}
        \end{equation*}
	}						
\end{algorithm}

\vspace{-2mm}

\subsection{Deployment (Online Execution) Procedure}
After convergence of the training procedure, the action-value function $Q_{\theta}$, with fixed parameters $\theta$, can be used to take actions to control different energy resources. To do this, the problem stated in \eqref{eq_optimal_action} must be solved. In this case, as function $Q_{\theta}$ represents a DNN, in order to solve \eqref{eq_optimal_action}, we leverage recent optimization results for DNNs. Thus, proposing a transformation of the DNN model $Q_{\theta}$ into a MIP formulation. 

\subsubsection{MIP for Deep Neural Networks}\label{sec:MIP_DNN}
\begin{figure}
   \centering
    \psfrag{A1}[][][1.0]{$s$}
    \psfrag{A2}[][][1.0]{$a$}
    \psfrag{A3}[][][1.0]{$Q(s,a)$}   
    \psfrag{\textit{\textbf{Q-value}}}[][][1.0]{$Q(s,a)$}    \includegraphics[width=0.5\columnwidth]{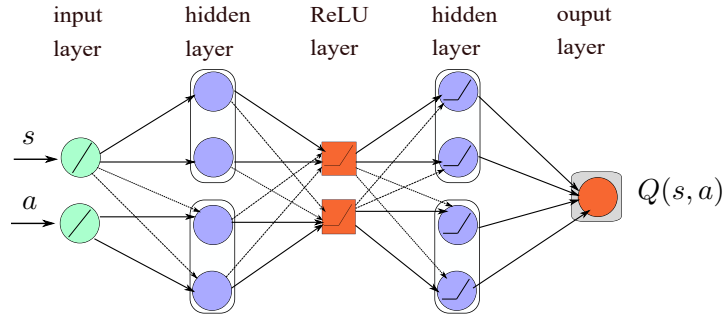}
   \caption{Layer structure of the DNN used to approximate the action-value function $Q(s,a)$. We denoted this DNN model as $Q_{\theta}(s,a)$ in Algorithm 1.}
   \label{fig_DNN_formulated Q-network}
\end{figure}

Let the DNN $Q_{\theta}(s,a)$ in Fig.~\ref{fig_DNN_formulated Q-network} consists of $K+1$ layers, listed from 0 to $K$. Layer 0 is the input of the DNN, while the last layer, $K$ refers to the outputs of the DNN. Each layer $k\in \{0,1,\dots,K\}$ have $U_k$ units, which is denoted by $u_{j,k}$, the $j_{th}$ unit of the layer $k$. Let $x^k$ refers to the output vector of layer $k$, then $x^k_{j}$ is the output of unit $u_{j,k},~(j=1,2,\dots,U_k)$. As layer 0 is the input of the DNN, then $x_j^0$ is $j_{th}$ input value for the DNN. For each layer $k\leq1$, the unit $u_{j,k}$ computes the output vector $x^k$ below: 

\begin{equation}\label{eq_general_x_output}
    x^{k}=h\left(W^{k-1} x^{k-1}+b^{k-1}\right)
\end{equation}

\noindent where $W^{k-1}$ and $b^{k-1}$ are matrices of weights and biases that compose the set of parameters $\theta$ i.e., $\theta=\{W,b\}$ and $h(\cdot)$ is the activation function, which in this case corresponds to the ReLU function, described as: for a real vector $y$, $\operatorname{ReLU}(y):=\max \{0, y\}$.

Based on the above definitions, the DNN of Fig.~\ref{fig_DNN_formulated Q-network}, with fixed parameters $\theta$, can be modeled as a valid MIP problem by modeling the ReLU function using binary constraints. Thus, using a binary activation variable $z^k_j$ for each unit $u_{j,k}$, the MIP formulation of a DNN can be expressed as~\cite{fischettiJo2018}: 

\begin{equation}\label{eq_goal}
\min_{\substack{x_j^k, s_j^k, z_j^k, \forall k}}  \left\{ \sum_{k=0}^{K} \sum_{j=1}^{l_{k}} c_{j}^{k} x_{j}^{k}+\sum_{k=1}^{K} \sum_{j=1}^{l_{k}} d_{j}^{k} z_{j}^{k} \right\}
\end{equation}
Subject to:
\begin{equation}\label{eq_relu_milp}
    \left.\begin{array}{r}
    \sum_{i=1}^{l_{k-1}} w_{i j}^{k-1} x_{i}^{k-1}+b_{j}^{k-1}=x_{j}^{k}-s_{j}^{k} \\
    x_{j}^{k}, s_{j}^{k} \geq 0 \\
    z_{j}^{k} \in\{0,1\} \\
    z_{j}^{k}=1 \rightarrow x_{j}^{k} \leq 0 \\
    z_{j}^{k}=0 \rightarrow s_{j}^{k} \leq 0
    \end{array}\right\} \\
    \forall k, \forall j,
\end{equation}

\vspace{-2mm}
\begin{equation}
\label{eq_bound_l0}
    l b_{j}^{0} \leq x_{j}^{0} \leq u b_{j}^{0}, \quad j \in l_{0},
\end{equation}

\vspace{-2mm}
\begin{equation}\label{eq_bound_k}
    \left.\begin{array}{l}
    l b_{j}^{k} \leq x_{j}^{k} \leq u b_{j}^{k} \\
    \overline{l b}_{j}^{k} \leq s_{j}^{k} \leq \overline{u b}_{j}^{k}
    \end{array}\right\} \forall k, \forall j.
\end{equation}

In the above formulation, weights $w_{i,j}^{k-1}$ and biases $b_j^{k}$ are fixed (constant) parameters; while the same holds for the objective function costs $c^k_j$ and  $d^k_j$. The ReLU function output for each unit is defined by \eqref{eq_relu_milp}, while \eqref{eq_bound_l0} and \eqref{eq_bound_k} define lower and upper bounds for the $x$ and $s$ variables: for the input layer ($k=0$), these bounds have physical meaning (same limits of the $Q_\theta$ inputs i.e., $s$ and $a$), while for $k\geq1$, these bounds can be defined based on the fixed parameters $\theta$~\cite{ceccon2022omlt}. Finally, notice that in order for the MIP formulation to be equivalent to the DNN, ReLU activation functions must be used, as explained in~\cite{fischettiJo2018}.

\subsubsection{Enforcing Constraints in Online Execution}
For an arbitrary state $s_t$, the optimal action $a_t$ can be obtained by solving the MIP in \eqref{eq_goal}--\eqref{eq_bound_k} derived from $Q_{\theta}$. In this case, as the decision variables are the actions $a_t$ (see~\eqref{eq_optimal_action}), the power balance constraint in \eqref{eq_balance} as well as the ramp-up and ramp-down constraints in \eqref{eq_ramping_up_constrain} and \eqref{eq_ramping_down_constrain}, respectively; can also be added to the MIP formulation described by \eqref{eq_goal}--\eqref{eq_bound_k}. As a result, the optimal actions obtained by solving this MIP \textit{strictly} enforce all operational constraints in the action space. This problem can be represented as,
\begin{equation}\label{eq_online_execution}
\begin{aligned}
 \max_{\substack{a\in {\cal A}, x_j^k, s_j^k, z_j^k, \forall k}} \quad & \left\{ \eqref{eq_goal} \right\}\\
\textrm{s.t.} \quad & \eqref{eq_relu_milp}-\eqref{eq_bound_k}, \eqref{eq_balance}, \eqref{eq_ramping_up_constrain}, \eqref{eq_ramping_down_constrain}.\\
\end{aligned}
\end{equation}

\begin{figure}[ht]
    \centering
    \psfrag{A1}[][][0.6]{$\text{Action space}$}
    \psfrag{A2}[][][0.6]{$\text{Feasible domain}$}
    \psfrag{A3}[][][1.0]{$a_1$}
    \psfrag{A4}[][][1.0]{$a_2$}
    \psfrag{A5}[][][1.0]{$a_3$}
    \psfrag{A6}[][][1.0]{$a_1$}
    \psfrag{A7}[][][1.0]{$a_2$}
    \psfrag{A8}[][][1.0]{$a_3$}
    \psfrag{x1}[][][1.0]{$t=t_0$}
    \psfrag{x2}[][][1.0]{$t=t_1$}
    \includegraphics[width=0.6\columnwidth]{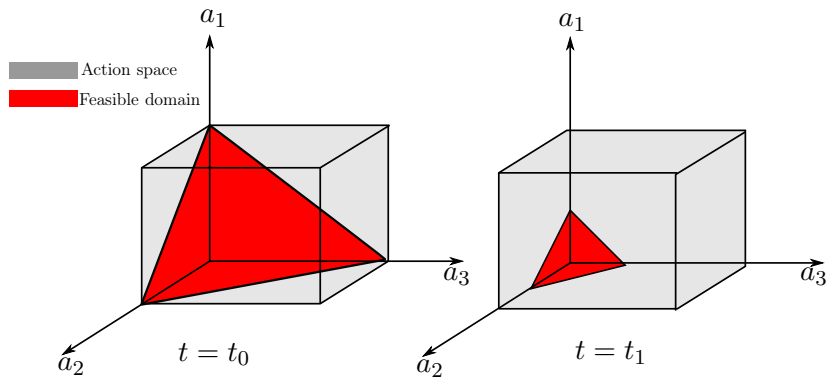}
    \caption{Action space (grey) and feasible action space (red) illustration. Actions  $a_1$, $a_2$, $a_3$ refer to generic actions in a three dimension action space ${\cal A}$. For each time step $t$, the power balance constraint in \eqref{eq_balance} can be seen as the hyperplane $a_1+ a_2+a_3=d$ that defines the feasible actions space.}
    \label{fig_action_space}
    \vspace{-2mm}
\end{figure}

\begin{figure}[ht]
   \centering
    \psfrag{A1}[][][1.0]{$h_{j}^{k}(\cdot)$}
    \psfrag{A2}[][][1.0]{$h_{j}^{k}(\cdot)$}
    \psfrag{A4}[][][0.8]{\textcolor{blue}{$a_1+a_2=d$}}
    \psfrag{A3}[][][1.0]{$Q_{\theta}(s,\cdot)$}       
    \psfrag{A5}[][][1.0]{$a_2$}
    \psfrag{A6}[][][1.0]{$a_1$} 
    \includegraphics[width=0.45\columnwidth]{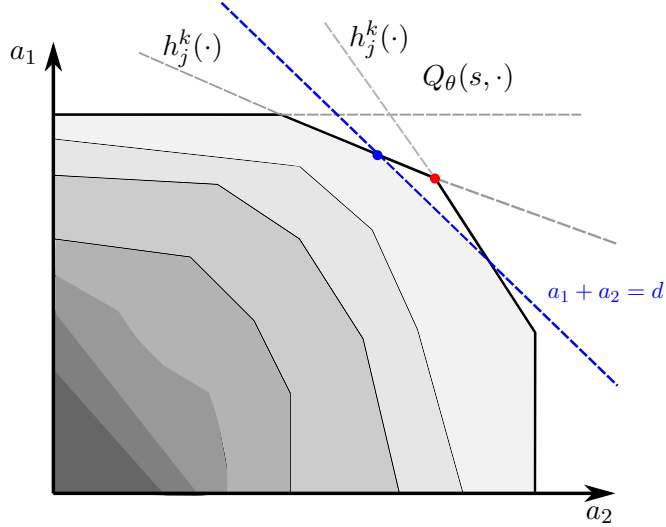}
   \caption {Visualization of the constraint space whose boundaries are formed by the hyperplanes $h_{j}^{k}(\cdot)$ defined by the ReLU activation functions derived from the deconstructed DNN $Q_{\theta}(s,\cdot)$ as a MIP formulation, for a specific state $s$ and actions $a_1$ and $a_2$. The grey are shows the increasing value (from darker to lighter) of $\nabla Q_{\theta}$.}
   The red point exemplifies the optimal solution of $\max_{\substack{a \in {\cal A}}} Q_{\theta}(s,\cdot)$ if constraint $a_1+a_2=d$ is disregarded. If such a constraint is added to the MIP formulation, the solution represented with the blue point will be reached.
\label{fig_DNN_formulated_MIP}
\end{figure}

\noindent To better understand the MIP formulation stated in~\eqref{eq_online_execution}, Fig.~\ref{fig_action_space} shows a re-interpretation of the power balance constraint in \eqref{eq_balance} as a hyperplane that define the feasibility region (for a three dimensional space) of the action space. Notice that such hyperplane may have different parameters for different time steps. Thus, if the hyperplane that enforces the power balance constraint is added to the MIP formulation that represents the DNN $Q_\theta$, the solution of such mathematical problem will ensure minimum operating cost (via the maximization of $Q_\theta$) and enforce all action space constraints, as exemplified in Fig.~\ref{fig_DNN_formulated_MIP}. In this case, this re-interpretation of the DNN as a MIP formulation offers enough flexibility to enforce equality constraints (as well as other constraints over the action space) for the energy system scheduling problem, such as the power balance. Algorithm~2 shows the step-by-step procedure used during the online execution of the proposed MIP-DQN algorithm.

\begin{algorithm}[ht]
	\caption{Online Execution for the MIP-DQN Algorithm}
	\label{algorithm2}
	Extract trained parameters ${\theta}$ from $Q_{\theta}$;\\
	Formulate the Q-function network $Q_{\theta}$ as a MIP formulation according to \eqref{eq_goal}-\eqref{eq_bound_k}. Add all action space constraints i.e., \eqref{eq_balance}, \eqref{eq_ramping_up_constrain} and \eqref{eq_ramping_down_constrain}.\\
	Extract initial state $s_0$ based on real-time data;\\
	\For{$t=1$ \KwTo $T$}
	{For state $s_t$, get optimal action by solving \eqref{eq_online_execution} using commercial MIP solvers;
	}						
\end{algorithm}

\section{Simulation Results and Discussions}
In this section, simulation results and discussions are presented. A comparison with DRL algorithms available in the literature, including PPO, SAC, DDPG and TD3 algorithms, is also presented. 

\vspace{-2mm}
\subsection{Case Study and Simulations Setup}
To test the developed MIP-DQN algorithm, an energy system consisting of three DG units and an ESS is defined. The DG unit's parameters are shown in Table~\ref{tab:dg_data_num}, while for the ESS, the charging/discharging limits, nominal capacity, and energy efficiency ($\eta_{B}$) are set to 100~kW, 500~kW, and 0.90, respectively. We assume that the network's maximum export/import limit is defined as 30 kW. To encourage the use of renewable energies, we set selling prices as half of the current electricity prices, i.e., $\beta=0.5$.

\begin{table}[ht]
	\centering
	\caption{DG units information} 
	\scalebox{0.8}{
		\begin{tabular}{cccccccc}
			\toprule
		Units & $a$[$\text{\$/kW}^2$] & $b$[\$/kW] & $c$[\$] & $\underline{P}^{G}$[kW] & $\overline{P}^{G}$[kW]& $RU$[kW]&$RD$[kW]\\
			\midrule
			$DG_1$ & 0.0034 & 3 & 30 & 10 & 150&100&100  \\
			\midrule
			$DG_2$ & 0.001 & 10 & 40 & 50 & 375&100&100  \\
			\midrule
			$DG_3$ & 0.001 & 15 & 70 & 100 & 500&200&200  \\
            \bottomrule
	\end{tabular}}%
	\label{tab:dg_data_num}%
	\vspace{-2mm}
\end{table}%

\begin{figure}[htbp]
    \centering
    \includegraphics[width=0.5\columnwidth]{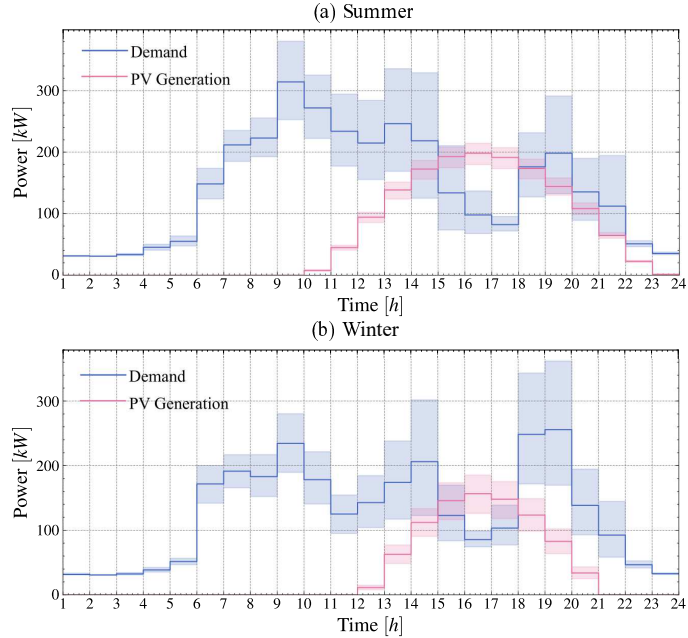}
    \caption{Mean and standard deviation of the demand consumption and PV generation.}
    \label{fig_demand_pv}
    \vspace{-2mm}
\end{figure}

One-year demand consumption and PV generation data are used as the original data-set, sampled in hour resolution. Fig.~\ref{fig_demand_pv} shows the mean and standard deviation of the demand consumption and PV generation during summer and winter for a period of 24h, defined as the length of one episode ($T=24$). The original dataset is divided into two additional datasets: training and testing. The training dataset contains the first three weeks of each month, while the testing dataset contains the remaining data. This allows the DRL algorithm to learn any seasonal and weekly behavior available in the PV generation and demand consumption data~\cite{shengren2022performance}. During training, the EES's initial SOC was randomly set. To implement our MIP-DQN algorithm, PyTorch and OMLT (see ~\cite{ceccon2022omlt}) package has been used. Default settings were used for all the implemented DRL algorithms, as shown in Table~\ref{tab:DRL_parameter}. All implemented algorithms are openly available in~\cite{Shengren}. Hyper-parameters $\sigma_1$ and $\sigma_2$ are defined as 0.01 and 20, respectively, as default values. Each test is run with five random seeds to eliminate randomness from code implementation.

\begin{table}[ht]
\centering
\caption{Parameters for DRL algorithms}
\scalebox{0.9}{
\begin{tabular}{cccccccc} 
\hline
Algorithm & Batch size $|B|$ & Learning rate & Buffer size $R$ & $\gamma$ & Network dimension & Optimizer  \\ 
\hline
DDPG       & 256        & 1e-4          & 5e4         & 0.995              & (64,64,64)        & Adam~      \\
SAC        & 256        & 1e-4          & 5e4         & 0.995              & (64,64,64)        & Adam~    \\
TD3        & 256        & 1e-4          & 5e4         & 0.995              & (64,64,64)        & Adam~       \\
PPO        & 256        & 1e-4          & -           & 0.995              & (64,64,64)        & Adam~       \\
MIP-DQN    & 256        & 1e-4          & 5e4         & 0.995             & (64,64,64)        & Adam~       \\ 
\hline
\end{tabular}}
\label{tab:DRL_parameter}
\end{table}

\subsection{Validation and Algorithms for Comparison}
In the research literature, DRL algorithms are usually compared with simple rule-based or MPC-based algorithms (considering the impacts of any forecasting error)~\cite{guo_optimal_2021}. Nevertheless, this procedure does not allow us to estimate the optimality gap between current DRL algorithms and the optimal global solution with a perfect forecast of the stochastic variables (i.e., generation and demand consumption). In this case, this optimal global solution with full knowledge should be regarded as an upper boundary, as none algorithm would perform better. Based on this, to validate and fairly compare the performance of the proposed MIP-DQN algorithm, besides comparing the optimal DERs schedule defined by several state-of-the-art DRL algorithms (DDPG, PPO, TD3), we compared with the optimal global solution obtained considering perfect forecast for the next 24 hours. In this case, the optimal global solution is found by solving the nonlinear mathematical programming formulation in Sec.~\ref{miqp_formulation}, implemented using Pyomo~\cite{hart_pyomo_2017}. Notice that different from the optimal global solution, all the tested DRL algorithms are able to make decisions only using current information. Finally, to evaluate the DRL algorithms' performance, the total operating cost, as in \eqref{eq_obj}, and the power unbalance, as in \eqref{eq_unb}, are used as metrics. 

\vspace{-2mm}
\subsection{Performance on the Training Set}\label{sec:training_small_case}
\begin{figure*}[ht]
    \centering
    \includegraphics[width=1.0\columnwidth]{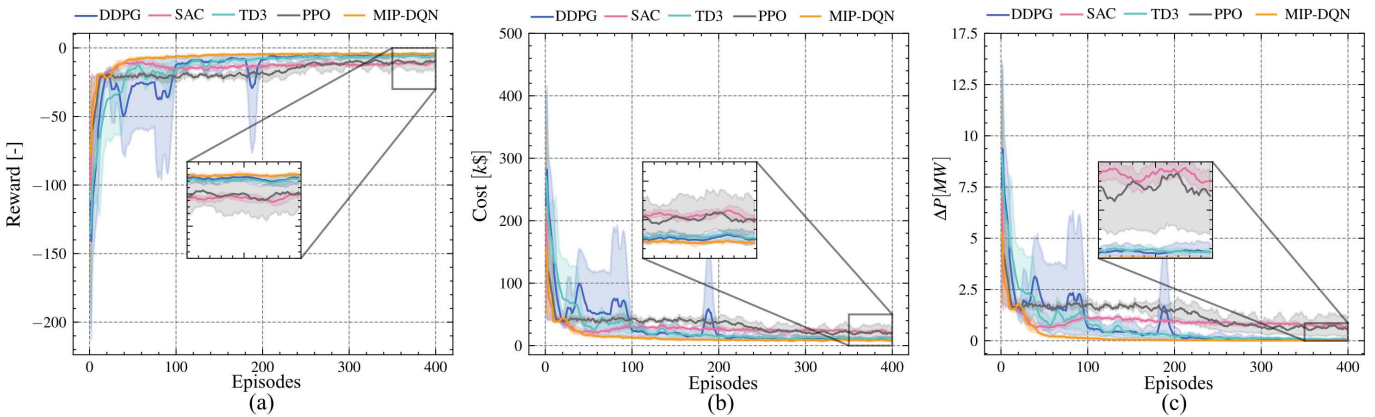}
    \caption{Mean and 95\% confident interval for the reward, operating cost and power unbalance for the developed MIP-DQN algorithm, as well as for other DRL algorithms, during training. As expected, none of these DRL algorithms are able to enforce the power balance constraint.}
    \label{fig_traing_process}
    \vspace{-2mm}
\end{figure*}

Figure~\ref{fig_traing_process} shows the average reward, operating cost, and power unbalance for the developed MIP-DQN algorithm and other DRL algorithms during the training process. As can be seen in Figure~\ref{fig_traing_process}, the average reward increases rapidly after 100 episodes of training, while the operating cost and the power unbalance significantly decrease. This behavior during training is typical of DRL algorithms as the DNN's parameters are randomly initialized, leading initially to random actions causing high power unbalance. Throughout the training, and due to the introduction of the penalty terms used in the reward definition in \eqref{eq_reward}, the DNN's parameters are updated, leading to higher quality actions, reducing power unbalance, and showing a lower operating cost. All algorithms converged before 400 episodes. After the last training episode, the power unbalance (presented by the average with 95\% confident interval) of DDPG, SAC, PPO, and TD3 are $64.8\pm99$ kW, $807\pm121$ kW, $65\pm18$ kW, $304\pm104$ kW, respectively; while a power unbalance of $12\pm15$~kW was observed for the proposed MIP-DQN algorithm. This result shows how the proposed MIP-DQN algorithm outperformed other DRL algorithms during the training process. Nevertheless, and as expected, none of the tested DRL algorithms (including the proposed MIP-DQN) can strictly enforce the power balance; if such algorithms are used in real-time operation, they might lead to unfeasible operation. Next, we show how our proposed algorithm can overcome this during online execution, even in unseen data.  

\vspace{-2mm}
\subsection{Performance on the Test Set}
\begin{figure}[t]
    \centering
    \includegraphics[width=0.55\columnwidth]{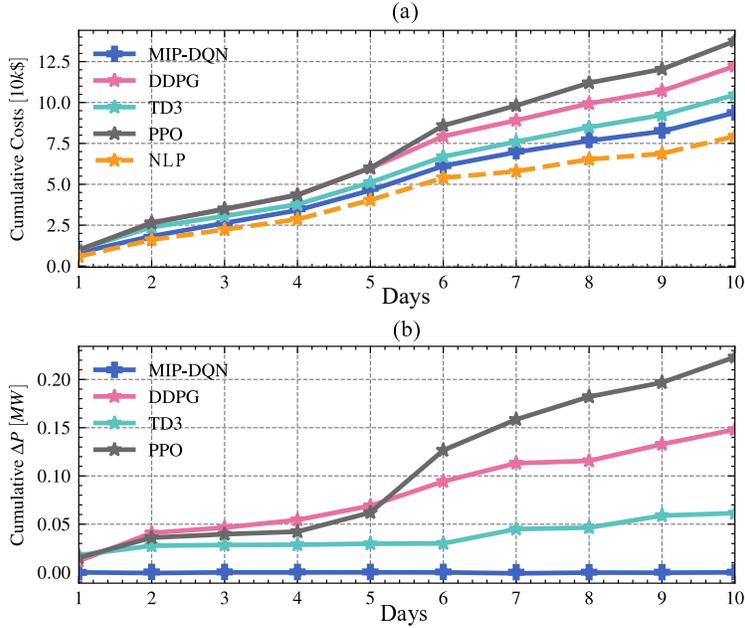}
    \caption{Cumulative costs and power unbalance for 10 days in the test set. The proposed MIP-DQN algorithm is able to strictly meet the power balance constraint while other DRL algorithms fail to do so.}
    \label{fig_cumulative_cost_unbalance}
    \vspace{-2mm}
\end{figure}

After training, the DNN's parameters of all the DRL algorithms are fixed as shown in Algorithm~2. A performance comparison is now made on the test set. Recall that the data on the test set is not used during training; therefore, it has not been seen by any of the DRL algorithms. To compare results on the test set, Fig.~\ref{fig_cumulative_cost_unbalance} shows the cumulative operating cost and power unbalance (which can be seen as a cumulative error) for 10 different days using the proposed MIP-DQN algorithm, as well as other DRL algorithms. The optimal global solution obtained by solving the NLP formulation and considering the perfect forecast is also presented. As can be seen in Fig.~\ref{fig_cumulative_cost_unbalance}, during online operation and for all 10 test days, the proposed MIP-DQN algorithm strictly meets the power balance constraint, while other DRL algorithms fail to deal with such equality constraint. Notice in Fig.~\ref{fig_cumulative_cost_unbalance} how DRL algorithms such as DDPG and TD3 reach a cumulative power unbalance near 0.14~MW at the end of the test period. As a result of such high unbalances, an operating cost of 53.3\% higher than the optimal global solution is also observed. In contrast, the proposed MIP-DQN algorithm achieves an operating cost of 94~$k\$$, i.e.,~17.6\% higher than the optimal solution.

To test the performance with a higher number of test days, Table~\ref{tab_compare_time} presents the average cumulative error (with respect to the solution obtained by solving the NLP formulation with perfect forecast), the average power unbalances, and total average computational time (over 30 test days) of the proposed MIP-DQN algorithm as well as other DRL algorithms. As can be seen, the proposed MIP-DQN algorithm has the lowest average error, 13.7\%; while strictly meeting the power balance (and other) constraint. In contrast, algorithms such as PPO showed poor performance reaching an error of 52.4\%. As expected, the total computational time required to execute the proposed MIP-DQN algorithm is higher than other DRL algorithms. This increase in the computational time is a result of the MIP formulation required to be solved in order to enforce the equality constraint (see \eqref{eq_online_execution}). Nevertheless, for this case, the proposed MIP-DQN algorithm can still be used for real-time operation as it only requires less than 20~s for execution. In this case, it is important to highlight that the computation time of the proposed MIP-DQN algorithm  is impacted by the size of formulated MIP problem, which is only determined by the size of the used Q network (layers, units of each layer, etc.) and not by the size of the energy system (microgrid) considered. Previous research has shown that (small) neural networks can generalize well in real environments~\cite{du2022deep,pinto2021data}, supporting the applicability of DRL models in real systems.

\begin{table*}[t]
\centering
\caption{Performance comparison of different DRL algorithms in a new test set of 30 days.}
\label{tab_compare_time}
\begin{tabular}{ccccl}
\cline{1-4}
\textbf{Algorithms} & \textbf{Error} & \textbf{$\Delta P~[MW]$} & \textbf{Computational time $[s]$ } &  \\ \cline{1-4}
\textbf{MIP-DQN} & \textbf{$13.7\mypm0.3\%$} & \textbf{0.0} & 17           &  \\
DDPG             & 47.3$\mypm$1.9\%          & 0.14$\mypm$0.021       & 4.3          &  \\
TD3              & 31.5$\mypm$0.7\%          & 0.06$\mypm$0.011       & 4.9 &  \\
PPO              & 52.4$\mypm$0.3\%          & 0.15$\mypm$0.007       & 4.3          &  \\ \cline{1-4}
\end{tabular}
\vspace{-2mm}
\end{table*}

\subsection{Dispatch Decisions Comparison}
\begin{figure}[]
    \centering
    \includegraphics[width=0.6\columnwidth]{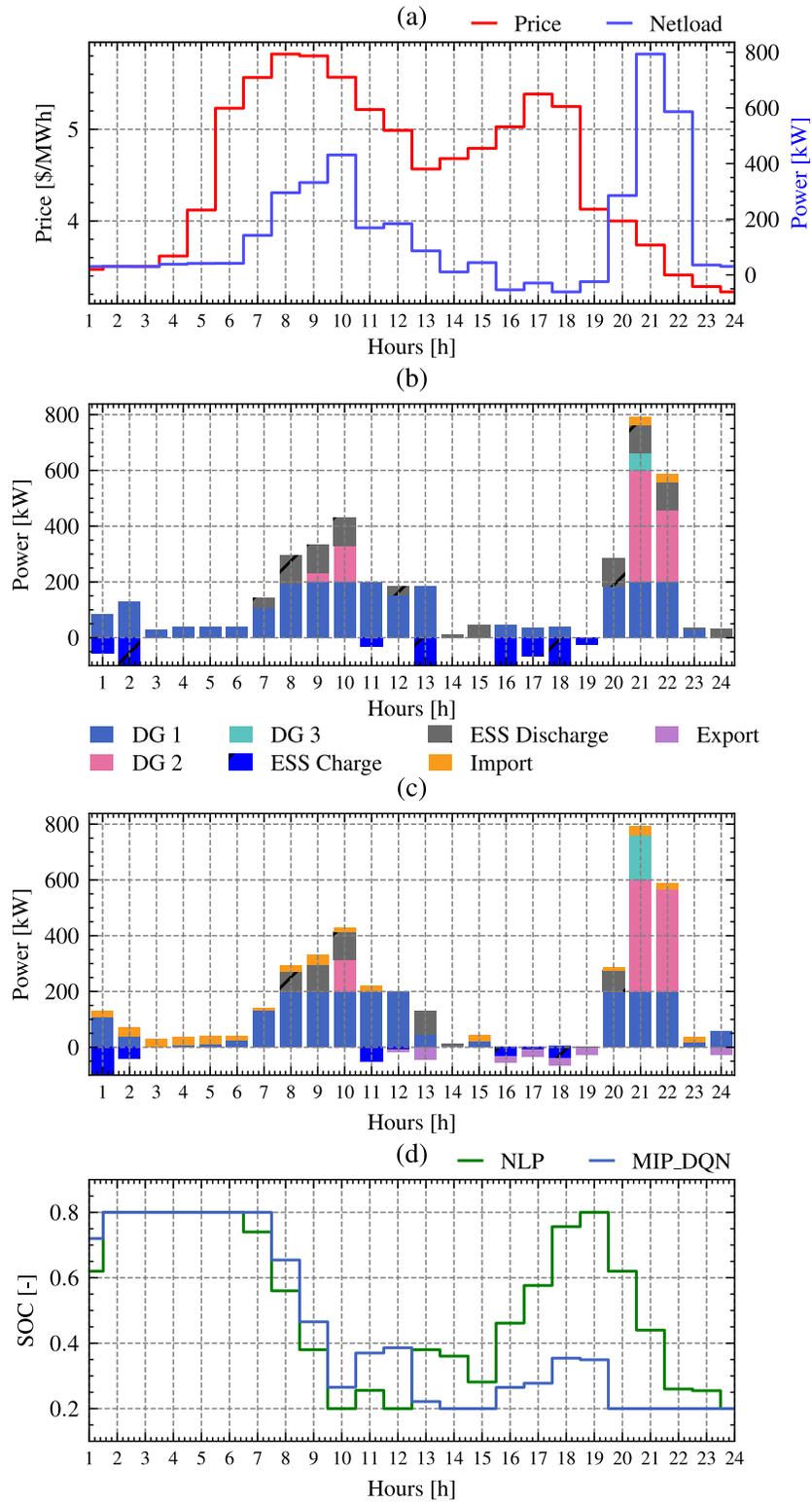}
    \caption{Operational schedule of all DG units and ESSs defined by the proposed MIP-DQN algorithm and the optimal global solution obtained by solving the NLP formulation considering perfect forecast.}
    \label{fig_detail_operation_schedule}
    \vspace{-2mm}
\end{figure}

Until now, the general performance of the proposed MIP-DQN algorithm has been presented, highlighting its capability of strictly enforcing the power balance constraint, even in unseen operational days. Next, a comparison in terms of the scheduling of the DG units and the ESSs is introduced. To do this, Fig.~\ref{fig_detail_operation_schedule} displays the output power of all the DG units, ESSs and the imported/exported power from the network for: the proposed MIP-DQN algorithm (Fig.~\ref{fig_detail_operation_schedule}$b$), and the optimal solution obtained after solving the NLP formulation considering perfect forecast (Fig.~\ref{fig_detail_operation_schedule}$c$).  Notice in Fig.~\ref{fig_detail_operation_schedule} that when the electricity price is high, and the net power is low, the proposed MIP-DQN algorithm dispatches the ESSs in charging mode, and a similar dispatch decision is observed in the optimal global solution. Notice also that, when compared with the optimal solution, the proposed MIP-DQN algorithm dispatched $3_{th}$ DG during the peak hour, which can be considered a sub-optimal decision as the operating cost of such DG is higher than the others.  This difference in this dispatch decision can be due to the estimated $Q$-function, which might not be good enough to represent the true action-value function. In this sense, as the proposed MIP-DQN algorithm chooses actions that maximize its $Q$-value estimation, the largest $Q$-value might not represent the best action for this specific state-action pair. Nevertheless, even in executing a sub-optimal decision, the proposed MIP-DQN algorithm is able to meet the power balance constraint, guaranteeing operational feasibility. Finally, although differences in the dispatch decisions made by the proposed MIP-DQN algorithm and the optimal solution can be observed, it is important to highlight that the optimal global solution is obtained considering the perfect forecast of the future generation and demand consumption for the next 24 hours, while the proposed MIP-DQN algorithm provides dispatch decisions in an hourly basis, without knowledge of the future values of the stochastic variables. 

\vspace{-2mm}
\subsection{Sensitivity Analysis}
\begin{figure}[h]
   \centering
   \includegraphics[width=0.5\columnwidth]{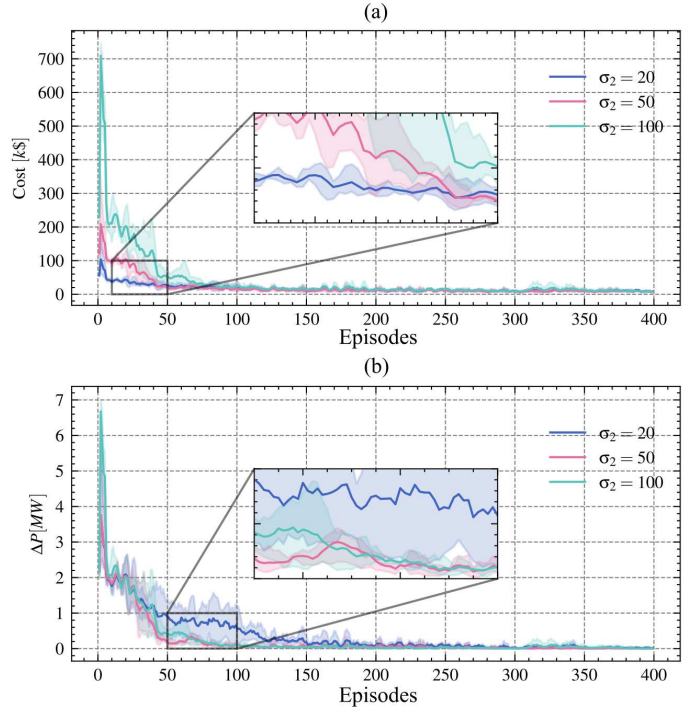}
   \caption{Average reward, operating cost, and power unbalance of the proposed MIP-DQN algorithm for different values of $\sigma_2$.}
   \label{fig_sensitive_analysis}
   \vspace{-2mm}
\end{figure}

To better understand the impact of hyperparameter $\sigma_2$ in the reward function in \eqref{eq_reward}, Fig.~\ref{fig_sensitive_analysis} shows the average operating cost and power unbalance (during training) for the proposed MIP-DQN algorithm for $\sigma_2=20, 50, 100$. As can be seen in Fig.~\ref{fig_sensitive_analysis}, and as expected, higher values of  $\sigma_2$ accelerate the convergence of the proposed MIP-DQN algorithm to rapidly reduce power unbalance, while having no apparent impact on the convergence of the operating cost. On the other hand, lower values of $\sigma_2$ seem to accelerate the convergence of the operating cost leaving behind the convergence of the power unbalance. In general, for the test performed, it was observed that the proposed MIP-DQN algorithm could converge in less than 200 episodes.     

\vspace{-2mm}
\subsection{Comparison with Safe DDPG Algorithm}

\begin{figure}[h]
   \centering
   \includegraphics[width=1.0\columnwidth]{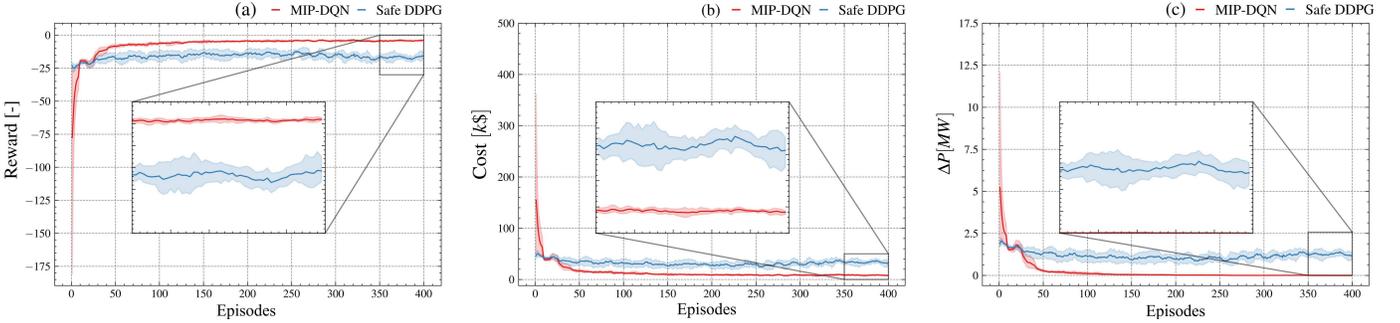}
   \caption{Mean and 95\% confident interval for the reward, operating cost and power unbalance for the developed MIP-DQN and Safe DDPG algorithms.}
   \label{fig_safe_ddpg_training}
   \vspace{-2mm}
\end{figure}

A comparison with current safe DRL algorithms is also performed. In this case, the proposed MIP-DQN algorithm is compared with a Safe DDPG algorithm, as presented in~\cite{safe_exploration_ddpg}. Fig.~\ref{fig_safe_ddpg_training} shows the average reward (Fig~\ref{fig_safe_ddpg_training}$a$), operating cost (Fig.~\ref{fig_safe_ddpg_training}$b$), and power unbalance (Fig.~\ref{fig_safe_ddpg_training}$c$) for the two algorithms being compared. In this case, and as expected, both algorithms fail to enforce the power unbalance constraint strictly during training. At the beginning of the training stage, the Safe DDPG algorithm shows a lower operating cost and power unbalance, and higher reward, when compared to the MIP-DQN algorithm. This is mainly due to the trained linear safe layer of the Safe DDPG, which projects the exploration action to a safer one, while the MIP-DQN algorithm is free to explore the action space regardless of the feasibility of the decided action. Nevertheless, along with the training, the Safe DDPG algorithm fails to learn to reduce further or eliminate power unbalance, while our proposed MIP-DQN algorithm reduces the unbalance sharply. This behavior is mainly due to the reward shaping of the MIP-DQN algorithm, which can learn to avoid the penalty due to the power unbalance during the training. It is important to highlight that the performance of the Safe DDPG algorithm depends on the quality of the trained safe layer that project the original action of the DDPG algorithm to a feasible one. In this case, as the safe layer is a linear function, its generalization capabilities may not be enough to learn the complex nonlinear energy system dynamic. Thus, even after projection, the action can not fully meet the power unbalance constraint. Moreover, as the safe layer modified the action during exploration, it also harms the performance of the trained RL algorithm as shown in Fig.~\ref{fig_safe_ddpg_training}. Compared to the Safe DDPG algorithm, the proposed MIP-DQN algorithm learns to eliminate the unbalance in a small value after training and guarantees the feasibility during the execution (Fig.~\ref{fig_cumulative_cost_unbalance}).

\vspace{-2mm}
\subsection{Larger Case Study}

\begin{figure}[h]
   \centering
    \includegraphics[width=1.0\columnwidth]{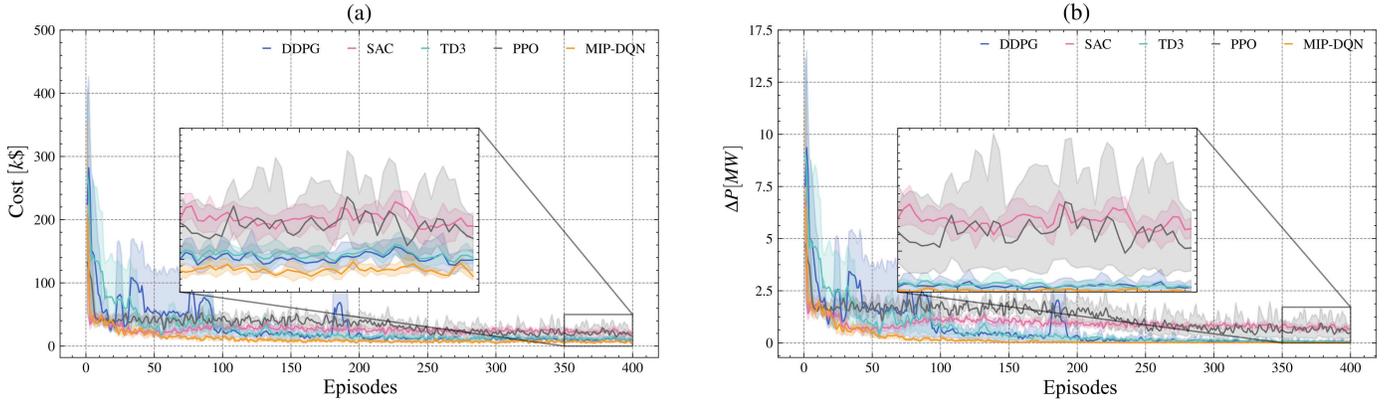}
   \caption{Mean and 95\% confident interval for the operating cost and power unbalance for the developed MIP-DQN algorithm, as well as for other DRL algorithms, during training.}
   \label{fig_more_battery_training}
   \vspace{-2mm}
\end{figure}

To test the performance of the proposed MIP-DQN algorithm on an energy system with multiple ESSs, an environment with three ESSs and three DG generators is designed. For this new environment, Fig.~\ref{fig_more_battery_training} shows the average operating cost and power unbalance of the proposed MIP-DQN algorithm as well as other state-of-the-art DRL algorithms, during the training process. As can be seen in Fig.~\ref{fig_more_battery_training}, the operating cost and power unbalance are significantly reduced. In this case, all tested DRL algorithms converged at around 400 episodes. The power unbalances (presented by the average with 95\% confident interval) of the DDPG, SAC, PPO and TD3 algorithms are $97\pm125$ kW, $533\pm208$ kW, $45\pm19$ kW, $462\pm98$ kW, respectively. In contrast, a power unbalance of $17\pm22$~kW was observed for the proposed MIP-DQN algorithm. Similar to the results presented in Sec.~\ref{sec:training_small_case} for the smaller case study (see Fig.~\ref{fig_traing_process}), none of the tested DRL algorithms can strictly enforce the power balance during training. Most of the observed power balance violations happen during peak load days, consistent with previous results~\cite{shengren2022performance}. Nevertheless, the proposed MIP-DQN algorithm is able to enforce power unbalance during the online execution, even on peak load days, as shown next. Additionally, compared to the result of simulations in Sec.~\ref{sec:training_small_case}, no performance degeneration is observed, proving the scalability of the proposed MIP-DQN algorithm.

\begin{figure}[h]
   \centering
   \includegraphics[width=1.0\columnwidth]{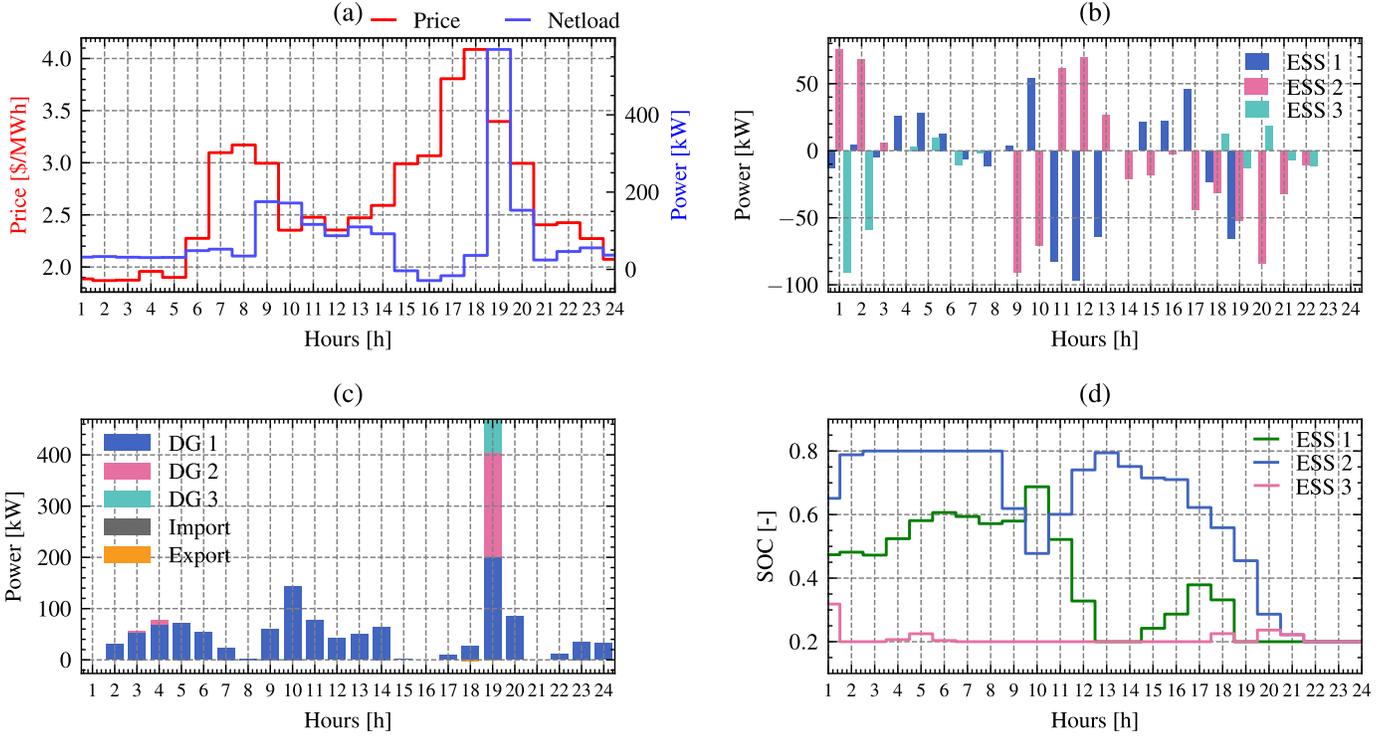}
   \caption{Operational schedule of all ESSs and DG units defined by the proposed MIP-DQN algorithm for a larger case study composed of three ESSs and three DG units.}
   \label{fig_more_battery_testing}
   \vspace{-2mm}
\end{figure}

Fig.\ref{fig_more_battery_testing} shows the scheduling decisions from the MIP-DQN algorithm for all three ESSs (Fig.\ref{fig_more_battery_testing}$b$) and DG generators (Fig.\ref{fig_more_battery_testing}$c$), and corresponding SOC changes (Fig.\ref{fig_more_battery_testing}$d$) in a typical day with extreme peak load. Notice that the power balance is strictly enforced during the peak load day. For instance, at 19h, the load is extremely high, and the MIP-DQN algorithm dispatches all the ESSs in discharging mode. This avoided importing electricity from the main grid as the electricity price was high at that particular time. These results showed that the proposed MIP-DQN algorithm learned to schedule feasible decisions for multiple ESSs in extreme peak situations. Notice also that, at hours 3 and 4, the proposed MIP-DQN algorithm dispatches the $2_{th}$ DG, instead of fully using the $1_{th}$ DG, which can be considered as a sub-optimal decision because the operating cost of $2_{th}$ DG is higher than that of $1_{th}$ DG. A similar result was observed in Fig.~\ref{fig_detail_operation_schedule}. Nevertheless, even in executing a sub-optimal decision, the proposed MIP-DQN algorithm is able to meet the power balance constraint, guaranteeing operational feasibility. Thus, the proposed MIP-DQN algorithm can provide feasible dispatch decisions hourly for multiple ESSs, displaying prominent scalability features.

\subsection{Discussion}
The penetration of renewable-based DERs energies significantly increases the uncertainty and complexity of the operation of energy systems. Existing model-based approaches may not perform well when defining the operational schedule of DERs in real time due to their poor accuracy and high computational time requirements. Due to this, current efforts are put into leveraging RL algorithms’ model-free and data-driven nature. After offline training, RL algorithms can provide near-optimal solutions in real-time. Nevertheless, the most critical challenge to enabling RL algorithms deployment in real energy systems scheduling frameworks is their lack of constraint enforcing guarantee. Even though several safe RL algorithms have tackled this problem, these approaches fail to meet the required security levels of energy systems operation~\cite{benchmark_SRL_review}. In general, model-based optimization approaches can guarantee the feasibility of the defined DERs schedule by setting hard constraints in the mathematical formulation, which is impossible to do in current RL algorithms.

To overcome the problem mentioned above, inspired by recent advances in deep learning and optimization research areas, we first bring constraint enforcement in RL algorithms combining deep learning and optimization theory. We developed a DRL algorithm, namely MIP-DQN, that can theoretically guarantee the feasibility of the decided solution and get the optimal solution during the online scheduling stage. To do this, we redesigned the training and online-scheduling procedure. The proposed MIP-DQN algorithm uses a trained $Q$-network to approximate the state-action values function. Exploration and exploitation are executed based on a trained policy network to update the Q-network parameters. After training, the $Q$-network is assumed to approximate the optimal $Q$-values. Then, the trained $Q$-network is extracted and formulated as MIP formulation, which can be used to impose hard constraints in the action space, ensuring the feasibility of the defined schedule. In this case, the power balance constraint is used as an example to show the effectiveness of the proposed approach. Results showed that MIP-DQN strictly meets the power balance constraint, showing a lower error when compared with other DRL algorithms and the optimal global solution. 

The essence of the proposed MIP-DQN algorithm is using a trained $Q$-network as a surrogate function for the optimal operational decisions. As above-mentioned, the optimality is defined by the $Q$-network modeled as a MIP formulation. Thus, the approximation quality of the $Q$ network determines the proposed algorithm's performance. In Fig~\ref{fig_detail_operation_schedule}, we showed that the proposed MIP-DQN could be considered a good quality operational schedule, albeit sub-optimal. Thus, efforts to reduce the error when compared with the optimal global solution must be centered on increasing the quality of the approximation of the Q-values via the used deep neural network. Additionally, the proposed MIP-DQN algorithm still needs to integrate a penalty term into the reward function to explore the right direction during the training process. This introduces extra hyperparameters that also impact the approximation performance of the obtained $Q$-function. An alternative exploration approach that can be used is to model the DNN as a MIP formulation in each iteration step; nevertheless, this would imply higher training time. 

\section{Conclusion}
This paper proposed a value-based DRL algorithm, namely MIP-DQN, to define the optimal dispatch decisions of multiple distributed energy resources within a renewable-based energy system. The proposed DRL algorithm was developed for continuous action (and state) spaces with the main feature of strictly enforcing all operational constraints in the action space during online execution, ensuring the feasibility of the defined schedule. This is done by re-formulating the deep neural network (DNN), used to approximate the action-value Q-function, as a mixed-integer programming (MIP) formulation enabling to further consider any action space constraint. Results showed that the proposed MIP-DQN algorithm obtained near-optimal solutions, with an error of 13.7\% when compared with the optimal solution obtained with a perfect forecast of the stochastic variables. A comparison with other DRL algorithms was also presented, observing higher errors than the proposed algorithm while failing to meet the power balance constraint on unseen test days. Future work directions include implementing \textit{plug-and-play} features, considering DERs' uncertain availability.
\bibliographystyle{elsarticle-num-names}
\bibliography{Ref} 
\vspace{-2mm}

\section{Appendix}
A sketch of a mathematical proof that ensures that the proposed MIP-DQN model provides the optimal solution while strictly enforcing linear constraints is presented below. To do this, we first assume the feasibility to the problem presented in Sec.~\ref{miqp_formulation} and also present (and adapt notation to match this paper) the Corollary 19, from \cite{serra2018bounding} as,

\textbf{Corollary 19:} If the input $(s,a)$ of the Q-network is a polytope and the DNN is a rectifier network (i.e., ReLU activation functions are used), then the mapping from input $(s,a)$ to the output $Q(s,a)$ of such a $Q$-network is mixed-integer representable.

The proof of Corollary 19 is available in \cite{serra2018bounding}. Note that this corollary implies that for any rectifier DNN, a mixed-integer formulation exists as long as the input is bounded. The Q-network used in the proposed MIP-DQN algorithm is a DNN with a rectifier activation function while the input $(s,a)$ are bounded as these correspond to the state and action variables as presented in Sec.~\ref{mdp_formulation}. We denote the optimal solution to this MIP formulation as $(s^{*},a^{*})$ whose optimal objective function value is $Q(s^{*},a^{*})$. 

Now, the extended MIP formulation obtained by adding on top of the MIP representation of the $Q(s,a)$ an equality constraint (in this case,~\eqref{eq_balance}) is also a feasible MIP representation. This is a consequence of the fact that such a mixed-integer representation of $Q(s,a)$ is composed of a set of linear regions whose unions form a bounded polyhedron (or polytope) (see Theorem 20 in \cite{serra2018bounding}), which we denote this here as ${\cal S}$ (see a representation in Fig~\ref{fig_DNN_formulated_MIP}). The addition of~\eqref{eq_balance} to ${\cal S}$, which is also a linear constraint, does not modify its nature of a bounded polyhedron (or polytope).

By exhaustion, two cases are distinguished: In the first case, the extended bounded polyhedron ${\cal S^{'}}={\cal S}\cup \eqref{eq_balance}$ is empty, rendering the solution of the MIP unfeasible, i.e., equality constraint in~\eqref{eq_balance} cannot be met. This is not possible as we assumed feasibility for the optimization problem. In the second case, ${\cal S^{'}}$ is not empty, in which an optimal solution exits and is feasible. If this is the case, and denoting such optimal solution as $(s^{'},a^{'})$, such solution meets the following condition: $Q(s^{'},a^{'}) \leq Q(s^{*},a^{*})$. This condition simply implies that $(s^{'},a^{'})$, by meeting the equality constraint in ~\eqref{eq_balance}, will at least have a q-value that is in the limit the same as the optimal solution $Q(s^{*},a^{*})$. This proves the fact that by solving the extended MIP formulation, a feasible and optimal solution that meets the equality constraint \eqref{eq_balance} is obtained. Nevertheless, it is important to highlight that optimality here relates to the good quality solution provided by the trained $Q$-network.

\end{document}